\documentclass[runningheads]{llncs}
\usepackage[utf8]{inputenc}

\usepackage{graphicx}

\usepackage[svgnames,dvipsnames]{xcolor}

\usepackage{hyperref}
\hypersetup{
  bookmarksopen=true,    
  pdftitle={Teaching Complex Systems based on Microservices},
  pdfauthor={Renato Cordeiro Ferreira},
  pdfcreator={pdflatex}, 
  pdfnewwindow=true,     
  colorlinks=true,       
  linkcolor=blue,        
  citecolor=DarkGreen,   
  filecolor=green,       
  urlcolor=DarkRed       
}

\definecolor{team-blue}{HTML}{307cdb} %
\definecolor{team-green}{HTML}{58bd89} %
\definecolor{team-orange}{HTML}{e69138} %
\definecolor{team-red}{HTML}{eb376b} %

\begin{document}
\title{
  Teaching Complex Systems based on Microservices
  \vspace{-5mm} 
}
\titlerunning{Teaching Complex Systems based on Microservices}
%
\author{
  Renato Cordeiro Ferreira\inst{1}\orcidID{0000-0001-7296-7091} \and \\
  Thatiane de Oliveira Rosa\inst{1,2}\orcidID{0000-0002-3980-0051} \and \\
  Alfredo Goldman\inst{1}\orcidID{0000-0001-5746-4154} \and \\
  Eduardo Guerra\inst{3}\orcidID{0000-0001-5555-3487}
}
\authorrunning{Ferreira et al.}
%
\institute{
  University of São Paulo, São Paulo, SP, Brazil\\
  \email{\{renatocf, thatiane, gold\}@ime.usp.br}\and
  Federal Institute of Tocantins, Paraíso do Tocantins, TO, Brazil\and
  Free University of Bozen-Bolzano, Bolzano, Italy\\
  \email{guerraem@gmail.com}
  \vspace{-5mm} 
}
\maketitle              
\begin{abstract}
Developing complex systems using microservices is a current challenge.
In this paper, we present our experience with teaching this subject to
more than 80 students at the University of São Paulo (USP), fostering
team work and simulating the industry's environment. We show it is
possible to teach such advanced concepts for senior undergraduate
students of Computer Science and related fields.
\end{abstract}

\section{Introduction}\label{sec:introduction}
\vspace{-1mm} 

The interest of industry and academia in the microservices architectural
style increases yearly. However, its adoption is non-trivial and has many
challenges. The teaching-learning process on this subject should cover
relevant technical and theoretical contents. It is important to think
about how universities can prepare students to develop complex systems
using microservices. Ideally, it should be interesting, motivating, and
offer an experience close to the industry.

This paper presents our approach for teaching the development of complex
systems based on microservices, as applied in the course \emph{``Laboratory
of Complex Computational Systems''} at the University of São Paulo (USP).
Since 2018, it has been offered four times as a two-week extension course,
with 65 students in total. In 2020, it was offered as a semester-long
course with 18 students.

Our main inspiration for this course was the XP Laboratory course%
~\cite{Goldman2019Agile}, also offered at USP. Another perspective
for teaching microservices was proposed by Lange et al.%
~\cite{Lange2019Microservices}, where students explored the subject
conceptually and then focused on strangling a monolithic application.

\section{About the Course}
\label{sec:about_course}
\vspace{-1mm} 

Our teaching method has three pillars: theoretical, technological
and practical. The first includes lectures about complex systems,
microservices and agile methodologies. The second is made of talks
about front-end and back-end Web development. The third is focused
on the implementation of an application based on microservices.

\newpage
In the course, lectures are given by researchers and industry professionals.
Students are organized into teams (4--6 members) and have to develop different
domains of the system. Our assessment is continuous and incremental. The final
grade is calculated based on presence and active participation in class, overtime
attendance during development sprints (four extra hours per week), and the
fulfillment of simple tests and exercise lists. The course also includes
regular warm-up activities to foster team building and to illustrate
concepts learned.

During the project development, we adopted agile practices such as
peer code review, pair programming, daily meetings, and sprint and
class retrospectives. To promote knowledge sharing and remote teamwork,
we used Discord and GitLab.

The total course duration was 120 hours (75 theoretical and 45 practical).
For a more detailed summary of the structure of the course, please access:
\href{https://web.archive.org/web/20200524194039/https://uspdigital.usp.br/jupiterweb/obterDisciplina?sgldis=MAC0475}{https://uspdigital.usp.br/jupiterweb/obterDisciplina?sgldis=MAC0475}.

\section{Course Project}
\label{sec:course_project}
\vspace{-2mm} 

As the course project, we presented \textsc{Hacknizer}: a platform to organize
and host hackathons. Some of the desired features for the system were: 
  user access control,
  hackathon creation and edition,
  sponsor and award registration,
  hackathon web page customization,
  participant registration,
  team composition,
  project submission, and
  winning project choice.

\begin{figure}[b]
  \vspace{-4mm} 
  \centering
  \begin{minipage}[b]{0.59\textwidth}
    \centering
    \includegraphics[width=\linewidth]{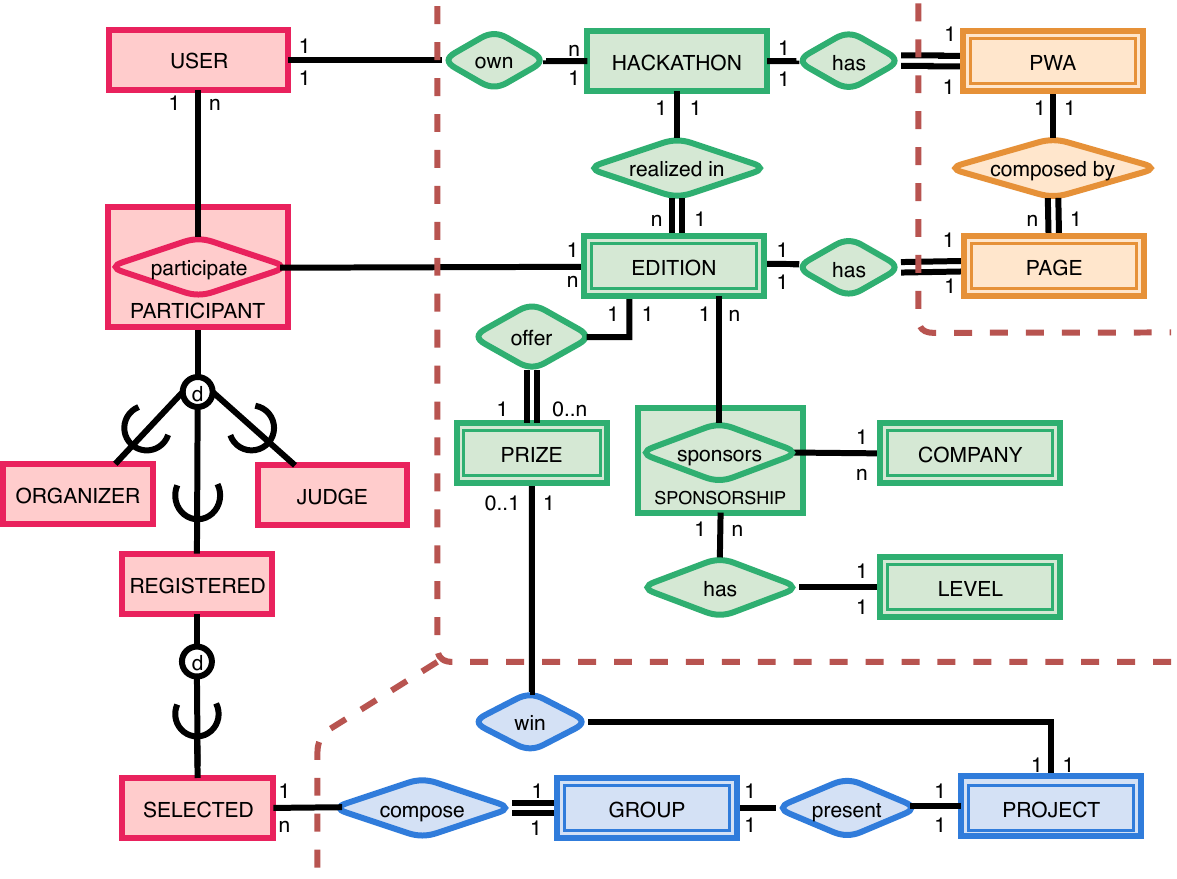}
    \caption{Entity–Relationship Model}
    \label{fig:mer}
  \end{minipage}%
  \hfill%
  \begin{minipage}[b]{0.39\textwidth}
    \centering
    \includegraphics[width=\linewidth]{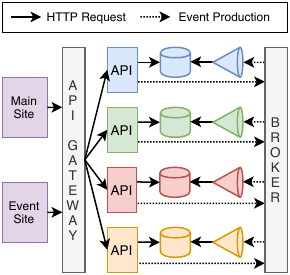}
    \caption{Architecture}
    \label{fig:architecture}
  \end{minipage}
\end{figure}

The 18 students were organized into four teams, composed of 4--5 members, with
different levels of knowledge and interests. We divided \textsc{Hacknizer} into
four bounded contexts:
  team management (\textbf{\textcolor{team-blue}{blue}}).
  hackathon management (\textbf{\textcolor{team-green}{green}}),
  user management (\textbf{\textcolor{team-red}{red}}), and
  web page customization (\textbf{\textcolor{team-orange}{yellow}}).
\autoref{fig:mer} presents an entity-relationship model that illustrates
the domain of the system.

We designed \textsc{Hacknizer} with a reactive microservice architectural
style~\cite{Boner2016ReactiveSystems} and adopted many patterns%
~\cite{Richardson2018}, such as
  \textsc{Service per Team},
  \textsc{Microservice Chassis},
  \textsc{API Gateway},
  \textsc{Single Database per Service},
  \textsc{Event Sourcing},
  \textsc{CQRS}, and
  \textsc{Saga}.
\autoref{fig:architecture} presents an architecture overview of the system.
\section{Lessons Learned}
\label{sec:lessons_learned}
\vspace{-1mm} 

\begin{figure}[b!]
  \centering
  \includegraphics[width=\linewidth]{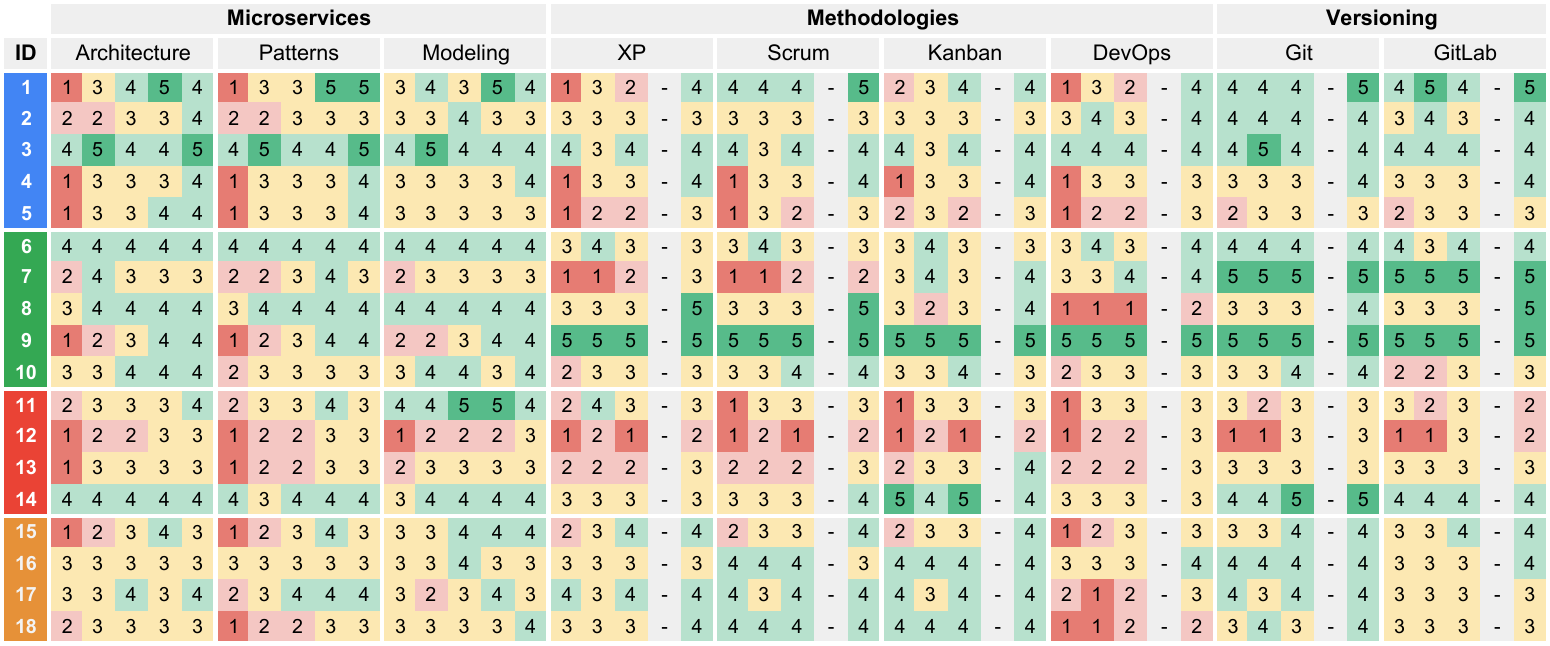}
  \includegraphics[width=\linewidth]{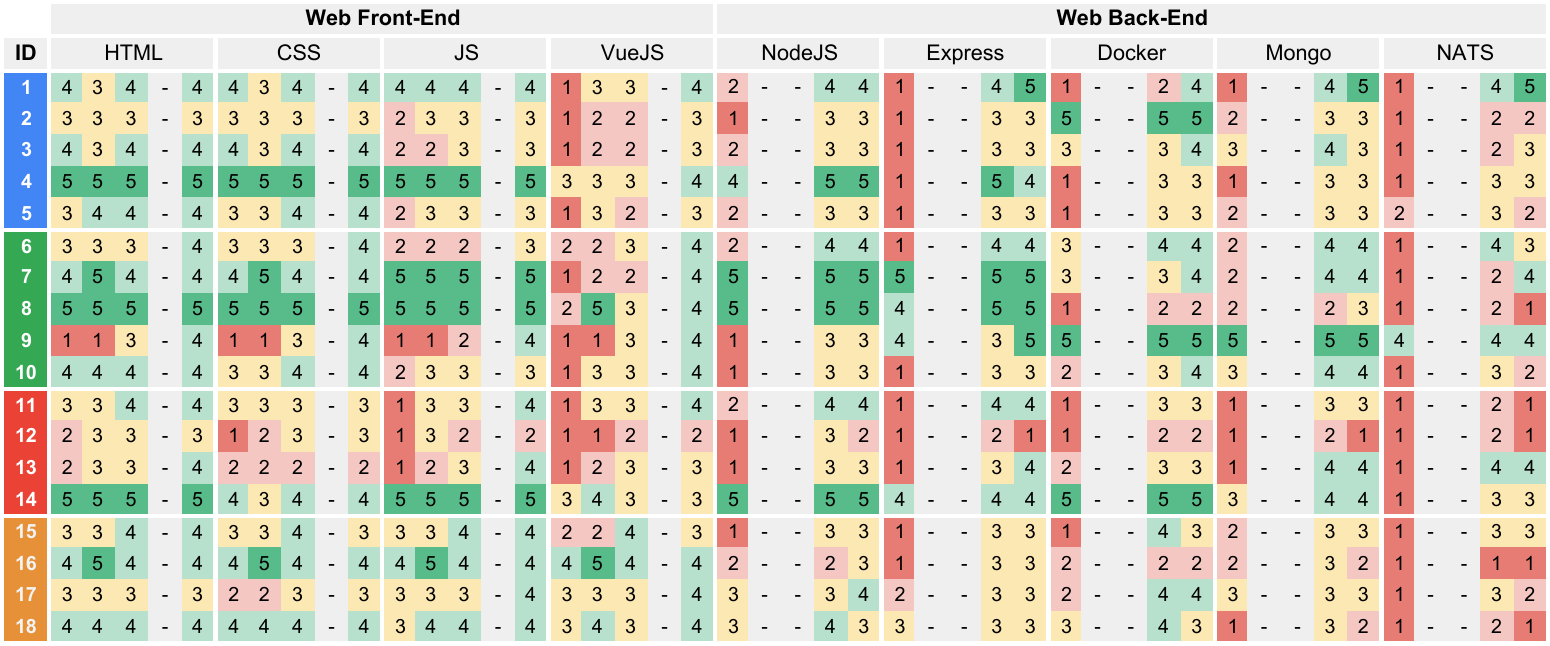}
  \vspace{-7mm}
  \caption{
    \textbf{Heatmap with students' self-assessed level of knowledge in 18
    skills the course aims to improve}. Each answer follows a Likert scale
    ranged 1-5, where 1 means ``very low'' and 5 means ``very high''. Each
    topic has answers about up to five moments: before the course, after
    the first lecture block, after the first sprint (both focused on theory
    and Web front-end), after the second lecture block, and after the second
    sprint (both focused on Web back-end). Dashes in the column represent
    a moment when the students were not asked about this subject.
  }
  \label{fig:heatmap}
\end{figure}

To evaluate the learning process, we surveyed our current 18 students
to collect their self-assessed level of knowledge in 18 skills the course
focuses on. Our results describe five different moments: before the course,
after the first lecture block, after the first sprint (both focused on theory
and Web front-end), after the second lecture block, and after the second sprint
(both focused on Web back-end). \autoref{fig:heatmap} shows a heatmap summarizing
the students' answers for the 18 skills and illustrates their evolution throughout
the course.

Regarding the microservices architectural style and its patterns, students
with low level of knowledge before the course ($S1$, $S4-S5$, $S9$, $S12-S13$
and $S15$) managed to evolve as the course progressed. On the other hand,
students who had a high level of knowledge ($S3$, $S6$ and $S14$) maintained it.
  
With respect to \emph{agile methodologies}, students who had attended the
XP Lab course~\cite{Goldman2019Agile} had a greater background
about these subjects. During our course, they either kept or evolved their
level of knowledge. Notwithstanding, all students with no previous experience
in agility advanced their knowledge.

About \emph{versioning}, students who are not enrolled in the Bachelor's
degree in Computer Science ($S5$, $S10$ and $S12$) had a low level of
knowledge in this subject before the course.

Regarding \emph{front-end technologies}, most students had experience
with the \mbox{basic} stack: HTML, CSS and JavaScript. The most unfamiliar
tool was \href{https://vuejs.org/}{Vue.js}, a modern single-page
application framework. Despite the challenges, even students with
very low level of knowledge ($S9$, $S11$ and $S13$) could get a
good understanding of how to develop with it.

Regarding \emph{back-end technologies}, students who had little experience
with Docker ($S1$, $S4$, $S5$, $S8$, $S10-S13$ and $S15-S17$) advanced
their knowledge. Two thirds of the students had a low level of knowledge about
\href{https://nodejs.org}{Node.js} and \href{https://expressjs.com}{Express},
but they assessed that they progressed to a medium or high level of knowledge
by the end of the course. Lastly, very few students knew how to use
\href{https://mongodb.com}{MongoDB} or \href{https://nats.io}{NATS}.
In the last survey, most of them had a good advance with the former,
but the same did not happen uniformly with the latter.


Besides the surveys, we scheduled a non-structured interview with the
students to ask eight questions about their backgrounds, difficulties
with each of pillar of teaching (theoretical, technological, practical),
and general impressions.

Overall, the course is meeting the students' initial expectations.
The greatest challenges reported include:
  team work, which is not commonly applied in other courses; and
  remote collaboration, since no single tool worked seamlessly
  for all students in all environments.
The main knowledge gains include:
  team work, because they felt they were learning how to develop together
  and were enjoying the collaborative discussions that came from it; and
  an environment similar to the industry's, since they reported our course
  was the closest to the challenges they expect to deal with in a full-time
  job, in particular the use of microservices.

\section*{Acknowledgment}
\label{sec:acknowledgement}
\vspace{-1mm} 

We want to thank our partner João Francisco Lino Daniel who helped to plan
and develop the course offerings described in this paper. We also thank
Alceste Scalas, Filipe Correia, and Rebecca Wirfs-Brock for suggestions
on how to improve this extended abstract.

%
%
%
 \bibliographystyle{splncs04}
 \bibliography{bibliography}

\end{document}